\documentclass[journal]{IEEEtran}

\usepackage{cite}

\ifCLASSINFOpdf
   \usepackage[pdftex]{graphicx}
   \graphicspath{{../pdf/}{../jpeg/}}
   \DeclareGraphicsExtensions{.pdf,.jpeg,.png}
\else
\fi
\UseRawInputEncoding
\usepackage{amsmath}
\usepackage{array}
\usepackage{standalone}
\usepackage{etoolbox}
\newtoggle{arXiv}
\newtoggle{bibrun}
\usepackage{newtxtext, newtxmath}
\usepackage{graphicx}
\usepackage{color, calc}
\usepackage{array, booktabs}
\newcolumntype{L}{>{$}l<{$}}
\newcolumntype{C}{>{$}c<{$}}
\newcolumntype{R}{>{$}r<{$}}
\usepackage{tikz, tikz-3dplot}
\usetikzlibrary{arrows, arrows.meta, calc, positioning}
\usetikzlibrary{decorations.pathmorphing}	
\tikzset{>=Stealth}
\usepackage{siunitx, physics}
\usepackage{enumitem}
\setlist[description]{labelindent=0pt, leftmargin=\parindent, font=\normalfont\itshape}
\usepackage{url}
\usepackage{subfigure}
\usepackage{textcomp}
\usepackage{eurosym}
\usepackage{xfrac}
\usepackage{pgfplots}
\usepackage{stmaryrd}
\pgfplotsset{compat=1.17}

\begin{document}
%
\title{Slice-selective Zero Echo Time imaging of ultra-short $T_2$ tissues based on spin-locking}

\author{\IEEEauthorblockN{
		Jos\'e\,Borreguero\IEEEauthorrefmark{2},
		Fernando\,Galve\IEEEauthorrefmark{1},
		Jos\'e\,M.\,Algar\'{\i}n\IEEEauthorrefmark{1},
		Jos\'e\,M.\,Benlloch\IEEEauthorrefmark{1}, and
		Joseba\,Alonso\IEEEauthorrefmark{1}}
	
	\IEEEauthorblockA{\IEEEauthorrefmark{1}MRILab, Institute for Molecular Imaging and Instrumentation (i3M), Spanish National Research Council (CSIC) and Universitat Polit\`ecnica de Val\`encia (UPV), 46022 Valencia, Spain}\\
	\IEEEauthorblockA{\IEEEauthorrefmark{2}Tesoro Imaging S.L., 46022 Valencia, Spain}\\%
	
\thanks{Corresponding author: F. Galve (fernando.galve@i3m.upv.es).}}

\markboth{Journal of \LaTeX\ Class Files,\,Vol.\,X, No.\,X, JANUARY\,2022}%
{Shell \MakeLowercase{\textit{et al.}}: Bare Demo of IEEEtran.cls for IEEE Journals}

\maketitle

\begin{abstract}
\newline
Purpose: {\normalfont To expand the capabilities of Zero Echo Time (ZTE) pulse sequences with a slice selection method suitable for the shortest-lived tissues in the body.}
\\
Methods: {\normalfont We introduce two new sequences that integrate spin-locking pulses into standard ZTE imaging to achieve slice selection: one for moderately short $T_2$ (DiSLoP), the other for ultra-short $T_2$ samples (PreSLoP). These methods exploit the slower signal decay (at $T_{1\rho}\gg T_2$) to retain the magnetization in the slices during the selection process, which is otherwise comparable to or even much longer than $T_2$.}
\\
Results: {\normalfont We demonstrate control over the slice profiles and positions for 2D imaging. We measure magnetization decay times during spin-locking ($T_{1\rho}$) as a function of pulse amplitude, showing significant lifetime enhancement for amplitudes as low as \SI{10}{\micro T}. We show imaging of slice-selected samples with $T_2$ characteristic times in the range of single milliseconds with DiSLoP and PreSLoP, and with the latter for sub-millisecond $T_2$ tissues. As compared to standard 3D ZTE sequences, PreSLoP achieves the same signal-to-noise ratio (SNR) in 2-5 times shorter scan times, and we argue that this is due to the filling scheme of the finite gap at the center of $k$-space unavoidable with ZTE sequences. Finally, we discuss a combination of DiSLoP with a dynamical decoupling sequence to avoid this central gap, leading to further scan time accelerations.}
\\
Conclusions: {\normalfont The proposed sequences are capable of slice-selected 2D imaging of tissues with $T_2$ as low as \SI{275}{\micro s} with good SNR within clinically acceptable scan times.}
\end{abstract}


 \ifCLASSOPTIONpeerreview
 \begin{center} \bfseries EDICS Category: 3-BBND \end{center}
 \fi
%
\IEEEpeerreviewmaketitle


\section{Introduction}

\IEEEPARstart{Z}{ero} Echo Time pulse sequences, otherwise known as Zero TE or ZTE sequences, are designed to capture the weak and fleeting signal emitted by hard biological tissues and solid-state matter in Magnetic Resonance Imaging (MRI) scanners \cite{ZTE1,ZTE2}. ZTE sequences are particularly suitable for ultra-short $T_2$ or $T_2^*$ MRI \cite{Weiger2011}, and have been successfully used for a wide variety of applications \cite{Weiger2019}, including imaging of tendons and bones \cite{Weiger2015,DeMello2020}, teeth \cite{Weiger2012,Algar_n_2020,Gonzalez2021}, myelin \cite{Seifert2017} or lungs \cite{Bae2020}. Two other families of sequences used for MR imaging of short $T_2$ samples are Ultra-short Echo Time (UTE, \cite{UTE1}) and SWeep Imaging with Fourier Transformation (SWIFT, \cite{Idiyatullin2006}). Despite their success, one limitation shared by all three families is that, in their basic forms, they are necessarily 3-dimensional, i.e. it is not possible to produce a 2D image of a pre-selected sample slice. 

The capability to selectively excite a given slice in the sample and obtain a 2D image is an essential part of the general MRI toolbox. With slice selection, the overall acquisition time for a single slice can be shortened, and 3D aliasing and ringing artifacts from slice to slice are avoided \cite{BkMcRobbie}. Besides, it can be critical for quantitative MRI, where furthermore slice profile assessment is a concern \cite{Keenan2019}.

Slice selection typically employs a soft radio-frequency (RF) pulse in the presence of a magnetic gradient perpendicular to the desired plane \cite{HOULT197969}. This long excitation stage is incompatible with the short-lived signals from hard tissues. Consequently, sequences for MR imaging of short $T_2$ tissues are inherently volumetric and the only slice selection procedure observed is unintentional and deleterious due to insufficient RF bandwidth to excite the complete Field of View (FoV). Slice selection can be realized with UTE by dividing the RF pulse into two halves, which are imaged sequentially and then merged into a unique signal \cite{Macovski2003}. Unfortunately, this doubles the scan time and the RF half pulses are sensitive to Eddy currents from the fast switching of the slice selecting gradient, making it challenging even with recent advances \cite{FABICH2014116, Stumpf2DUTE}. For ZTE and SWIFT sequences, we are not aware of any prior work illustrating the possibility of slice selection.

In this paper, we demonstrate a technique for slice-selective ZTE of ultra-short $T_2$ species, based on quantum spin-locking techniques in the presence of a linear magnetic gradient field \cite{Wind_1978}. We show that spin-locking is a versatile, robust and well suited complement to ZTE (and potentially other) pulse sequences. During spin-locking, a single-tone RF excitation of amplitude $B_\text{1SL}$ is resonant only with spins in the selected slice, due to the presence of the gradient. This ``locks'' the selected magnetization to the transverse direction, and slows down the signal loss characteristic time from $T_2$ to $T_{1\rho}$ \cite{callaghan}, which can be an order of magnitude longer than $T_2$ for hard tissues (where $T_1\gg T_2$). In the following sections, we show control over the position, shape and thickness of the selected slice, and present 2D and 3D ZTE images of ultra-short $T_2$ samples and hard tissues including cow-bone or teeth. We call the simpler sequence DiSLoP (\textit{Direct Spin-Locked PETRA}), where imaging can take place right after slice selection for moderately short $T_2$ times. Optionally, we can include a preservation pulse \cite{ROMMEL1990264} after spin-locking to safely store the magnetization while the slice and encoding gradients are switched off/on, respectively, allowing for ultra-short $T_2^*$ 2D-MRI. We call this variant PreSLoP (\textit{Preserved Spin-Locked PETRA}), which we find to be clean, immune to Eddy currents, and delivers a higher signal-to-noise ratio (SNR) per unit time than both DiSLoP and PETRA. PETRA stands for Pointwise Encoding Time-reduction with Radial Acquisition and is a well-known ZTE sequence where the unavoidable gap at the center of $k$-space, typical of these acquisitions, is filled in a pointwise fashion \cite{Grodzki2012,Algar_n_2020}.


\begin{figure*}
	\centering
	\includegraphics[width= 2\columnwidth]{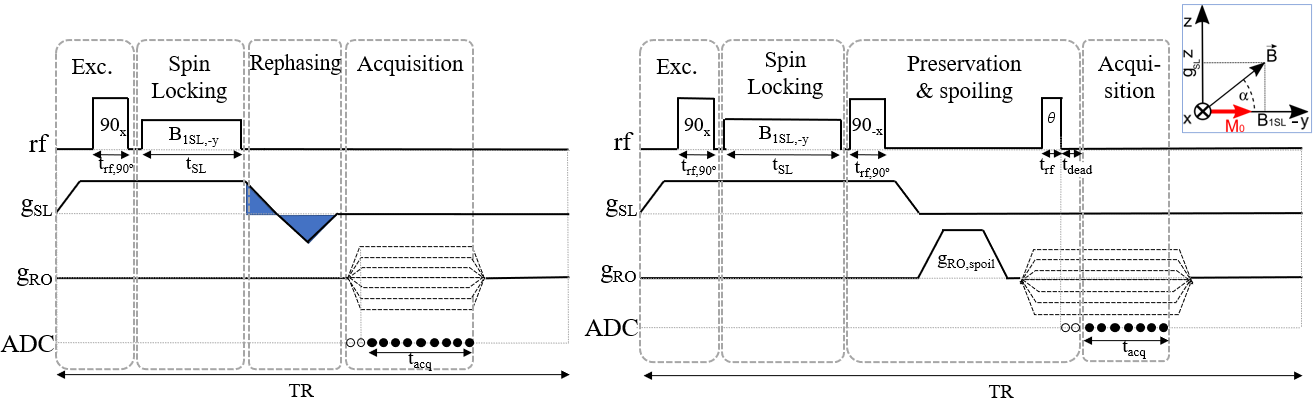}
	\caption{Scheme of proposed protocols for 2D imaging of short $T_{2}$ samples. Left) DiSLoP. Right) PreSLoP. Inset) Rotating frame field vectors. The empty circles in the ADC (analog-to-digital converter) line indicate the dead time before the start of the acquisition, which in turn leads to the gap at the center of $k$-space.}
	\label{fig:pulseseqdiagram}
\end{figure*}

\section{Theory}

\subsection{Spin-locking}
\label{sec:spinlocking}
 
Spin-locking (SL) was reported in the 1950s by A.\,G.\,Redfield while investigating Nuclear Magnetic Resonance saturation (NMR) in solids \cite{Redfield1955}, applied in the late 1970s as an alternative means of slice selection by R.\,A.\,Wind \emph{et al.} \cite{Wind_1978}, it is now a mainstay in the field of solid-state NMR \cite{BkSchmidt-Rohr}, and is broadly used in MRI for $T_{1\rho}$-weighted imaging \cite{Borthakur}. In the context of this work, its main advantage compared to the usual soft pulse approach for slice selection is that in-slice magnetization is kept aligned (locked) with the resonant RF field. Consequently, the transverse magnetization is lost at a rate defined by $T_{1\rho}$, rather than $T_2$. In general, $T_{1\rho}$ is significantly longer than $T_{2}$ and its dependence on $B_\text{1SL}$ is weak. This enables the detection of short-lived tissues and samples whose signal would have faded away after slice selection by common procedures, where spectrally selective RF pulses are typically $>1$\,ms.

Slice selection through spin-locking works by first applying a gradient $g_\text{SL}$ in the slice direction ($z$ in the inset of  Fig.\,\ref{fig:pulseseqdiagram}), and then a hard $90^{\circ}$ pulse which flips all sample magnetization into the plane transversal to $B_{0}$. Immediately after this, the RF locking field is pulsed on for a time $t_\text{SL}$. This has strength $B_\text{1SL}$, it is phase-shifted by $90^{\circ}$ with respect to the first excitation pulse so that it is aligned with the precessing magnetization direction ($-y$ in Fig.\,\ref{fig:pulseseqdiagram}), and it is resonant with $\gamma \left( B_{0} + g_\text{SL} z_{0} \right)$, with $z_{0}$ the position of the slice to be selected. In the inset of Fig.\,\ref{fig:pulseseqdiagram} we show the situation for $z_0=0$: spins which are close to this plane see only a locking field along $-y$, while off-slice spins mostly see a field $g_\text{SL}\cdot z$ along $z$, with a rather abrupt transition between both regimes. Therefore, off-slice magnetization is both homospoiled by the inhomogeneity created by $g_\text{SL}$ and also dephased by its intrinsic short $T_{2\rho}$ (transverse decay in the rotating frame), while {\it in-slice} magnetization is locked by $B_\text{1SL}$ and decays with the much longer $T_{1\rho}$. The selected slice has an approximate width (see Appendix)
\begin{equation}\label{eq:thickness}
 \Delta z \approx 2B_\text{1SL}/g_\text{SL}.
\end{equation}
The spin-locking time required for the slice to be selected, which we arbitrarily define as having suppressed off-slice magnetization contributions by around 90\,\% in an homogeneous sample, can be roughly estimated as
\begin{equation}\label{eq:time}
	t_\text{SL} \gtrsim \frac{7\pi}{2\Omega_\text{1SL}}\,\text{, where } \Omega_\text{1SL} = \gamma B_\text{1SL} = \gamma g_\text{SL} \Delta z/2,
\end{equation}
with $\gamma$ the gyromagnetic factor ($\approx 2\pi\cdot 42$\,MHz/T for protons). Slice selection can be faster for short $T_{2\rho}$ samples. The detailed equations governing magnetization are derived in the Appendix. All in all, the experimental knobs for slice selection are $g_\text{SL}$, $B_\text{1SL}$ and $t_\text{SL}$. In the rotating frame, the longitudinal relaxation time becomes $T_{1\rho}$, which is given by the spectral density of heat-bath fluctuations at the frequency $\gamma \sqrt{B_\text{1SL}^2+(g_\text{SL} \cdot z)^2}$. Thus the fundamental condition for SL to work is that
\begin{equation}
	T_2^*\ll t_\text{SL}\ll T_{1\rho},
\end{equation}
where $T_2^*$ in this case can be extremely short because it is due to $g_\text{SL}$.

\subsection{Advanced spin-locking}
As in standard slice selection, the shape of the slice achieved with SL is determined by the Fourier Transform of the time envelope of the resonant RF pulse. The diagrams in Fig.\,\ref{fig:pulseseqdiagram} show square pulses, which result in spatial profiles following a sinc funtion. It is sometimes convenient to produce sharp rectangular profiles, e.g. for quantitative MRI, so that off-slice contributions do not contaminate the resulting reconstructions \cite{Josan2009}. Slice selection based on SL also allows for this, using a single-tone field with a sinc modulation envelope. Although we demonstrate the viability of this approach in Fig.\,\ref{fig:SquareSL}, we encounter two disadvantages with respect to square SL pulses: i) a sharp square profile necessitates several sinc lobes, thus lengthening the spin-locking duration, and ii) the modulation of $B_\text{1SL}(t)$ means $T_{1\rho}$ is given by a sampling of the heat-bath spectral function at locking fields lower than $B_\text{1SL}^\text{max}$, thus leading to a shorter `effective' $T_{1\rho}$, which is the crux of the matter. For this reason, we have used unmodulated locking fields in the remaining studies.

A further interesting possibility is multi-slice selection, which can speed up imaging significantly through the simultaneous excitation and readout of multiple slices, either through RF phase encoding, gradient phase encoding or by parallel imaging coil arrays (see \cite{BreuerSMS} for a review). This advantage can be translated to slice selection based on spin-locking by means of a multi-tone excitation resonant with spins at multiple locations. Although we have not explored this experimentally in this work due to hardware limitations, we refer to the Appendix for a more detailed discussion.

\section{Methods}
In the following we present and motivate DiSLoP and PreSLoP (Fig.\,\ref{fig:pulseseqdiagram}), the slice-selective Zero Echo Time pulse sequences we have devised for 2D imaging of short $T_{2}$ samples. All experiments have been performed in a 260\,mT custom-built DentMRI- Gen I system described elsewhere \cite{Algar_n_2020}. Images are reconstructed with Algebraic Reconstruction Techniques (ART)\cite{kaczmarz,Gordon1970}, which do not require density compensation and allow for highly undersampled $k$-space data \cite{Algar_n_2020}. $T_1$ characteristic times have been measured by inversion recovery \cite{Bydder1998} and $T_2$ times by spin-echo \cite{Carr1958,Meiboom1958}, both by fitting single or double exponential functions (see Tab.\,\ref{tab:SampleRelaxationParameters}). Magnetic field inhomogeneities can be shimmed down to the 10\,ppm level with small gradient corrections in the 260\,mT system. This corresponds to $T'_2\sim 10$\,ms, so $T_2$ and $T_2^*$ values are very similar for all samples used  in this work, with the sole exception of ham, since magnetic susceptibilities do not play a significant role. Where given, SNR values are estimated by averaging in three different image regions with and without sample (see Appendix and Fig.\,\ref{fig:ROIs}).

\subsection{DiSLoP: Direct Spin-Locked PETRA}
This sequence is designed for 2D imaging of samples whose $T_2$ are not extremely short, in the order of units of milliseconds. DiSLoP starts with an \textit{excitation block}, where we ramp up the slice selection gradient $g_\text{SL}$ and later transfer the whole sample magnetization to $-y$ with a hard $90^{\circ}_x$ RF pulse, i.e. a pulse that rotates the magnetization by $90^{\circ}$ around the $x$ axis in the frame of reference that revolves at the spin precession frequency. Importantly, this pulses needs to be short and intense enough to span the bandwidth given by $g_\text{SL}$ in the sample volume. Then comes the \textit{spin-locking block}, where we switch on $B_\text{1SL}$ along $-y$ for a time $t_\text{SL}$ long enough to dephase off-slice spins. This time can be estimated from Eq.\,(\ref{eq:time}) or found empirically if the sample consists mostly of short $T_{2\rho}$ components. For a given slice width, slice selection is faster with higher $g_\text{SL}$ values, which in turn requires also higher $B_\text{1SL}$ strength, as seen from Eqs.\,(\ref{eq:thickness}) and (\ref{eq:time}). This is an important criterion if one seeks to retain in-slice signal coherence, which decays as $T_{1\rho}$ during SL. Third comes the \textit{rephasing block}, where we ramp down $g_\text{SL}$ as fast as possible to constrain the $T^{*}_{2}$ decay of in-slice spins. An extra gradient blip - previously calibrated to prevent distortions due to Eddy currents - compensates the in-slice dephasing caused by ramping down $g_\text{SL}$. Finally, in the \textit{acquisition block} we acquire data along radial spokes in $k$-space following the ZTE procedure, i.e. we ramp up a magnetic gradient $g_\text{RO}$ along the readout direction, and we start the acquisition after its onset. Alternatively, one could start acquiring data while ramping up the gradient, as in UTE, but this would have precluded the performance comparisons we present in Sec.\,\ref{sec:Results}.

\subsection{PreSLoP: Preserved Spin-Locked PETRA}
The PreSLoP protocol is similar to DiSLoP, but it is designed to prevent $T_{2}^*$ decay by replacing the rephasing block with a \textit{preservation \& spoiling block} prior to acquisition. This is achieved by a preservation pulse (hard $90^\circ_{-x}$ RF pulse) directly after spin-locking, which places the in-slice magnetization along $z$. In this way we preserve the coherence while we switch off/on the slice/encoding gradients, which is crucial for 2D imaging of tissues with extremely short $T_2$. During this switching, an additional spoiling gradient pulse can be included to remove remaining off-slice coherences, even if we have found this unnecessary in our system. In the acquisition block of PreSLoP, the magnetization can be excited to an arbitrary flip angle $\theta$ and the acquisition starts after the onset of $g_\text{RO}$, as in a standard PETRA sequence. This is an advantage compared to DiSLoP, since the long scan times inherent to PETRA can be partly compensated by using Ernst angle excitations and shortening the repetition time TR \cite{Weiger2011}. For simplicity, however, we have used $\theta=90^\circ$ throughout the paper.

PreSLoP is subject to distortions in the slice profile due to Eddy currents for short repetition times (TR). Inverting the polarity of $g_\text{SL}$ for each radial acquisition mitigates this effect, because the contributions average out to a large extent. All the PreSLoP images in this paper are therefore taken in this way (see Appendix and Fig.\,\ref{fig:polarity}).

\subsection{DiSLoP vs PreSLoP}
DiSLoP is a simpler sequence than PreSLoP and potentially faster because it can be made immune to the finite dead time ($t_{\text{dead}}$) required to switch the electronics from transmission (Tx) to reception (Rx) mode, thus avoiding the Cartesian single-point encoding scheme to fill the $k$-space gap in PETRA. Instead, in DiSLoP one can simply ramp up and down the encoding gradient $g_\text{RO}$ during the rephasing block, so that acquisitions start at $k=0$ and pointwise encoding is unnecessary. Unfortunately, this possibility requires accurate control and calibration of Eddy currents for different values of $g_\text{RO}$ in each repetition, so in the results below we have used the simpler DiSLoP sequence shown in Fig.\,\ref{fig:pulseseqdiagram}, which includes a pointwise-encoding stage.

The main advantage sought with PreSLoP is its (partial) immunity to $T_{2}^*$ decay after the spin-locking block, allowing for much higher SNR images than DiSLoP for hard biological tissues, solid-state matter, and samples with short-lived MR signals in general. To quantify this gain, in the following sections we show images of the same sample with both sequences under conditions as similar as possible.

\subsection{1D and 3D imaging}
DiSLoP and PreSLoP are conceived as 2D imaging sequences where the through-slice spin contribution is integrated in the recorded signals. Nevertheless, we show also images taken with sequence variants where the slice selection gradient is used for encoding as well. In this way, a 3D image of the slice-selected sample is obtained, which can be useful to benchmark the efficiency of the slice selection block. We call these 3D-DiSLoP and 3D-PreSLoP for clarity. We also find it useful to run the 1D versions of our sequences for fast characterization of slice profiles of simple, homogeneous samples (see Figs.\,\ref{fig:SLcontrol}-\ref{fig:ChainPulses} in Sec.\,\ref{sec:Results}). 

\section{Results \& analysis}
\label{sec:Results}

\subsection{Control over spin-locking and preservation blocks}\label{sec:control}

\begin{table}[h]
	\caption{Relaxation parameters of employed samples and figures where they have been used.}
	\centering
	\begin{tabular}{c c c c}
		\hline
		Parameter & $T_1$ (ms) & $T_2$ (ms) & Figures \\
		\hline
		York ham  & \SI{289}{ms}  & \SI{18}{ms} & \ref{fig:t1rhosamples} \\
		1\% CuSO$_{4}$ doped water & \SI{16}{ms}  & \SI{7.5}{ms} & \ref{fig:t1rhosamples}  \\
		3\% CuSO$_{4}$ doped water & \SI{3}{ms}  & \SI{2.5}{ms} & \ref{fig:SLcontrol},\ref{fig:SquareSL},\ref{fig:ChainPulses},\ref{fig:t1rhosamples},\ref{fig:3DDiSLoPcontrol}  \\
		Honey & \SI{14.4}{ms}  & \SI{1.5}{ms} & \ref{fig:t1rhosamples},\ref{fig:honeysample}  \\
		Eraser  & \SI{27}{ms}  & \SI{800}{\micro s} & \ref{fig:t1rhosamples}  \\
		Photopolymer resin  & \SI{23.1}{ms}  & \SI{650}{\micro s}  & \ref{fig:t1rhosamples} \\
		Clay  & \SI{36}{ms}  & \SI{550}{\micro s} & \ref{fig:t1rhosamples},\ref{fig:clayphantom}  \\
		Bone  & \SI{7}/\SI{30}{ms}  & 315 / \SI{630}{\micro s} & \ref{fig:t1rhosamples},\ref{fig:bone}  \\
		Teeth  & \SI{3}/\SI{29}{ms}  & \SI{275}{\micro s} / \SI{6}{ms} & \ref{fig:t1rhosamples},\ref{fig:horsetooth}  \\
		\hline
	\end{tabular}
	\label{tab:SampleRelaxationParameters}
\end{table}

\begin{figure*}[t]
	\includegraphics[width= 2\columnwidth]{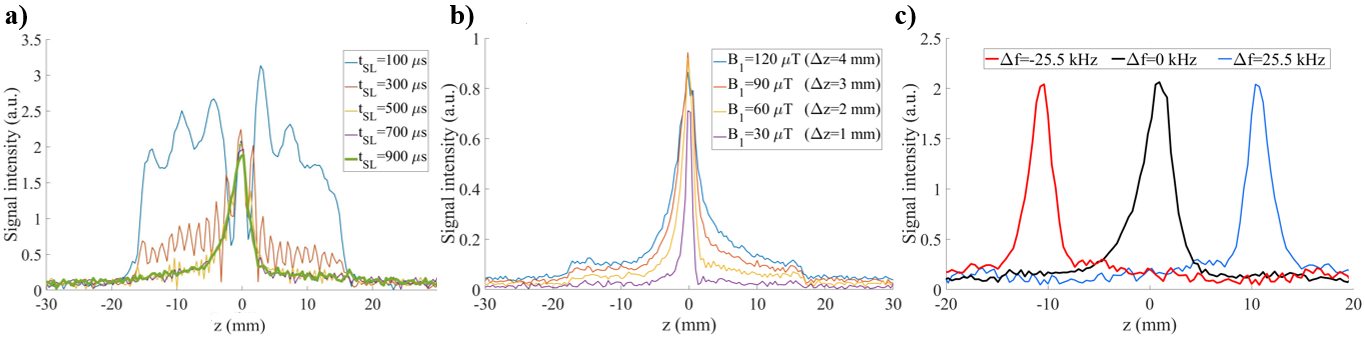}
	\caption{Control over the spin-locking block with 1D-DiSLoP and a PLA (polylactic acid) cuboid container filled with 3\,\% CuSO$_{4}$ doped water. a) Slice profile as a function of $t_\text{SL}$ ($B_\text{1SL} \approx \SI{90}{\micro T}$, $g_\text{SL} \approx 60$\,mT/m). b) Slice thickness as a function of $B_\text{1SL}$ ($t_\text{SL}= \SI{900}{\micro s}$, $g_\text{SL} \approx 60$\,mT/m). c) Slice position as a function of Larmor detuning ($B_\text{1SL} \approx \SI{90}{\micro T}$, $g_\text{SL} \approx 60$\,mT/m, $t_\text{SL}= \SI{900}{\micro s}$).}
	\label{fig:SLcontrol}
\end{figure*}

In this section we first demonstrate experimental control over the slice selection process with 1D-DiSLoP and a cuboid container filled with 3\,\% CuSO$_{4}$ doped water ($T_1$ and $T_2$ in Tab.\,\ref{tab:SampleRelaxationParameters}). The hard $90^\circ$ RF pulses are typically \SI{10}{\micro s} long with an amplitude $\approx \SI{550}{\micro T}$. We focus initially on the \emph{spin-locking time} required for slice selection. Here, after the slice selection block, we use a rephasing block and then lift an encoding gradient in the direction of slice selection. In Fig.\,\ref{fig:SLcontrol}a we show the resulting 1D profiles after different SL times, and observe that slice selection is practically achieved for $t_\text{SL} \gtrsim \SI{500}{\micro s}$ with $B_\text{1SL}\approx\SI{90}{\micro T}$ and $g_\text{SL}\approx 60$\,mT/m, for $\Delta z \approx 3$\,mm. From Eq.\,(\ref{eq:time}) we would expect the SL procedure to need $\approx \SI{460}{\micro s}$. We have observed that the estimation in Eq.\,(\ref{eq:time}) suffices in the general case, even if for \textit{ex vivo} hard biological tissues we have used shorter times because their short $T_{2\rho}$ values speed up slice selection. In this sense we found it useful to reproduce Fig.\,\ref{fig:SLcontrol}a for each type of sample, to find the shortest $t_\text{SL}$ in each case, but always checking the image obtained by 3D-PreSLoP for correct slice selection.

To demonstrate control over the \emph{slice thickness} with 1D-DiSLoP, we fix $t_\text{SL}= \SI{900}{\micro s}$ and $g_\text{SL} \approx 60$\,mT/m and use multiple spin-locking field amplitudes ($B_\text{1SL} \approx 30,\, 60,\, 90,\, \SI{120}{\micro T}$, corresponding to slice thicknesses $\Delta z \approx 1,\,2,\,3,\,4$\,mm, see Fig.\,\ref{fig:SLcontrol}b). Since the slice-selected magnetization follows a Lorentzian profile (see Appendix), the off-slice contribution is more notable for higher $\Delta z$. Therefore, square SL pulses are more convenient for thinner slices, and thicker slices may require sinc-modulated pulses. The profile asymmetries observed are compatible with a decaying drift of $B_0$ caused by Eddy currents. We have observed these in many other scenarios and for the images presented in this paper we suppress their effect substantially with interleaved data acquisition, where each radial spoke uses a different $g_\text{SL}$ polarity.

Finally, to show control over the \emph{slice position}, we set $B_\text{1SL} \approx  \SI{90}{\micro T}$, $g_\text{SL} \approx 60$\,mT/m and $t_\text{SL}= \SI{900}{\micro s}$ ($\Delta z \approx 3$\,mm), and we detune the SL frequency by $\Delta f=-25.5$, 0 and 25.5\,kHz from the Larmor frequency at the gradient isocenter, shifting the selected slice to $z_0 \approx -10$, 0 and 10\,mm respectively. 

\begin{figure}
	\centering
	\includegraphics[width=1\columnwidth]{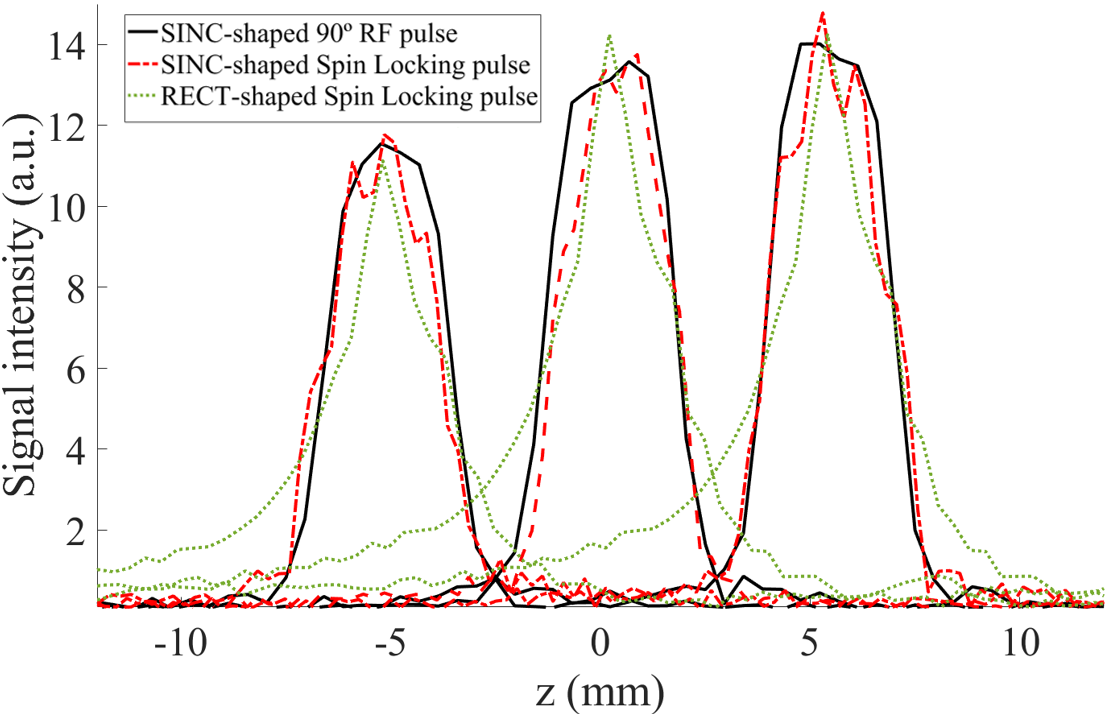}
	\caption{Comparison between square (green) and sinc-modulated spin-locking (red) and sinc-modulated standard slice selection (black) for the same 3\,\% CuSO$_{4}$ doped water sample as in Fig.\,\ref{fig:SLcontrol}. We have normalized all curves to the same height for ease of comparison (see text).}
	\label{fig:SquareSL}
\end{figure}

Next we investigate the performance of square and sinc-modulated SL pulses and the slice profiles they produce (Fig.\,\ref{fig:SquareSL}). For this we use the same sample and experimental parameters as in Fig.\,\ref{fig:SLcontrol}c. Because its $T_2$ is not extremely short (see Tab.\,\ref{tab:SampleRelaxationParameters}), we could slice-select also by means of a standard sinc-modulated excitation using gradient echo encoding (i.e. without spin-locking), with the shortest echo time possible in our setup. Both sinc functions have five lobes in total. We observe similar profile transitions for both sinc-modulated experiments, and a peaked profile for spin-locking without modulation, with longer tails, as expected. We have normalized the profiles in Fig.\,\ref{fig:SquareSL} to the same peak values for visual aid. However, the standard gradient echo acquisition has a lower signal than the constant SL curve by a factor of 1.5, since the magnetization is lost at a rate given by $T_2^*$ instead of $T_{1\rho}$. The signal is even lower with sinc-modulated SL (factor of 3.2), due to the longer SL pulse duration and the shorter effective $T_{1\rho}$.

\begin{figure}
	\centering
	\includegraphics[width=1\columnwidth]{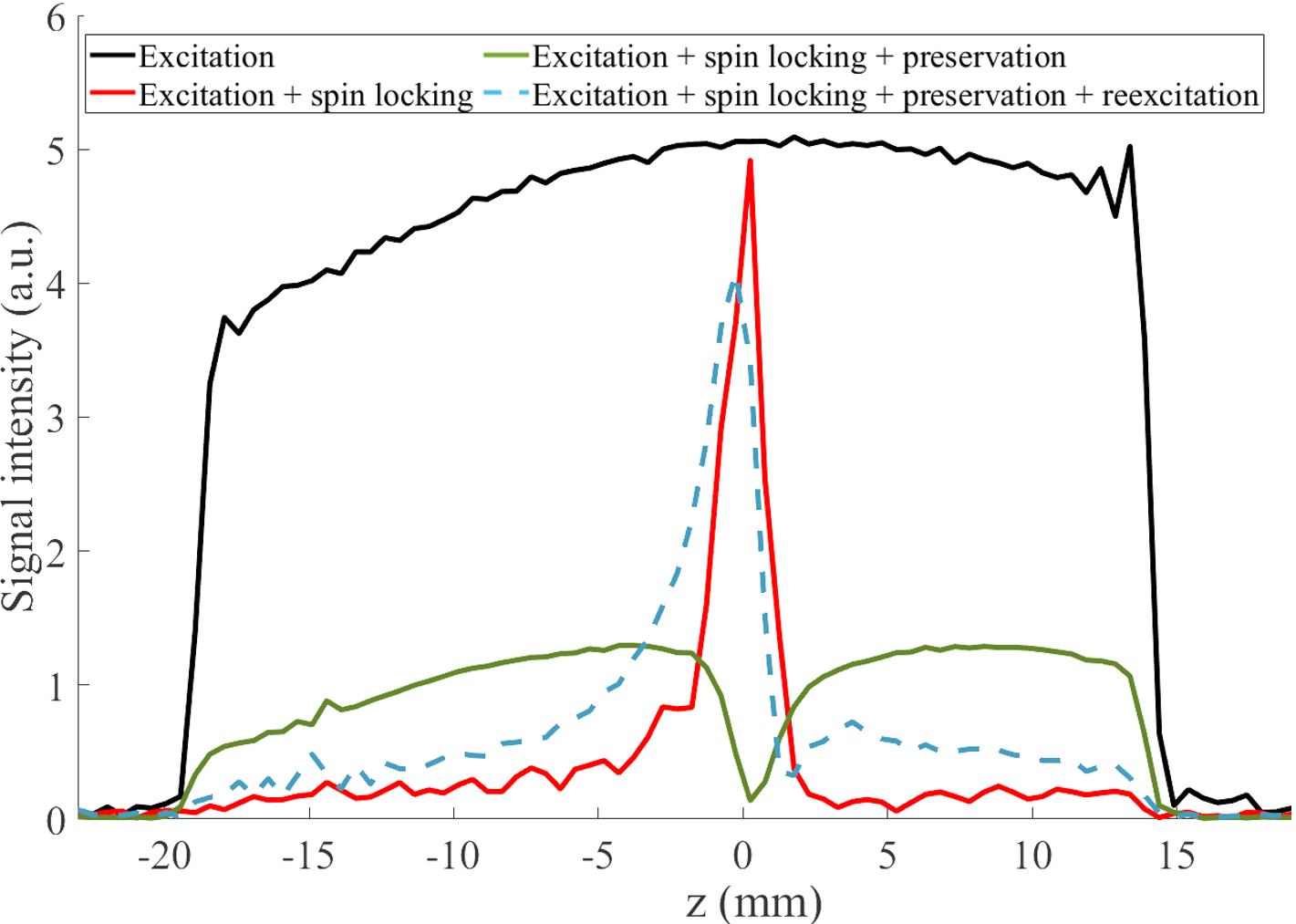}
	\caption{Profile of the transverse magnetization after each stage in PreSLoP for the same 3\,\% CuSO$_{4}$ doped water sample as in Figs.\,\ref{fig:SLcontrol} and \ref{fig:SquareSL}, with $\Delta z \approx2$\,mm: after initial $90^\circ_x$ excitation (black), after spin-locking (red), after $90^\circ_{-x}$ storage pulse (green), and after final $90^\circ_x$ excitation (dashed blue). In the last case, there is a delay $\tau=\SI{200}{\micro s}$ where gradients are ramped between the $90^\circ_{-x}$ and $90^\circ_{x}$ pulses. We have used $B_\text{1SL} \approx \SI{60}{\micro T}$, $g_\text{SL} \approx 60$\,mT/m, $t_\text{SL}=1$\,ms. The full (black) curve is not flatter due to capillarity effects and bubble formation in the small PLA container.}
	\label{fig:ChainPulses}
\end{figure}

Having characterized our control with DiSLoP, we now move to PreSLoP. Figure\,\ref{fig:ChainPulses} shows the evolution of the 1D sample profile (3\,\% CuSO$_{4}$ solution) along the sequence: after the initial $90^\circ$ excitation, after spin-locking, after the preservation pulse (-$90^\circ$) and after the final excitation pulse (prior to image encoding). For each of these stages we 1D-encode the transverse magnetization after the chosen set of steps. We have used $B_\text{1SL} \approx \SI{60}{\micro T}$, $g_\text{SL} \approx 60$\,mT/m, $t_\text{SL}=1$\,ms and $\Delta z\approx 2$\,mm. After the preservation pulse, only off-slice magnetization is observed, since in-slice magnetization has been stored along $z$ and its transverse component is negligible. The profiles after preservation and in the final stage are asymmetric presumably due to drifting Eddy currents, as in Fig.\,\ref{fig:SLcontrol}b.

\begin{figure}
	\centering
	\includegraphics[width=1\columnwidth]{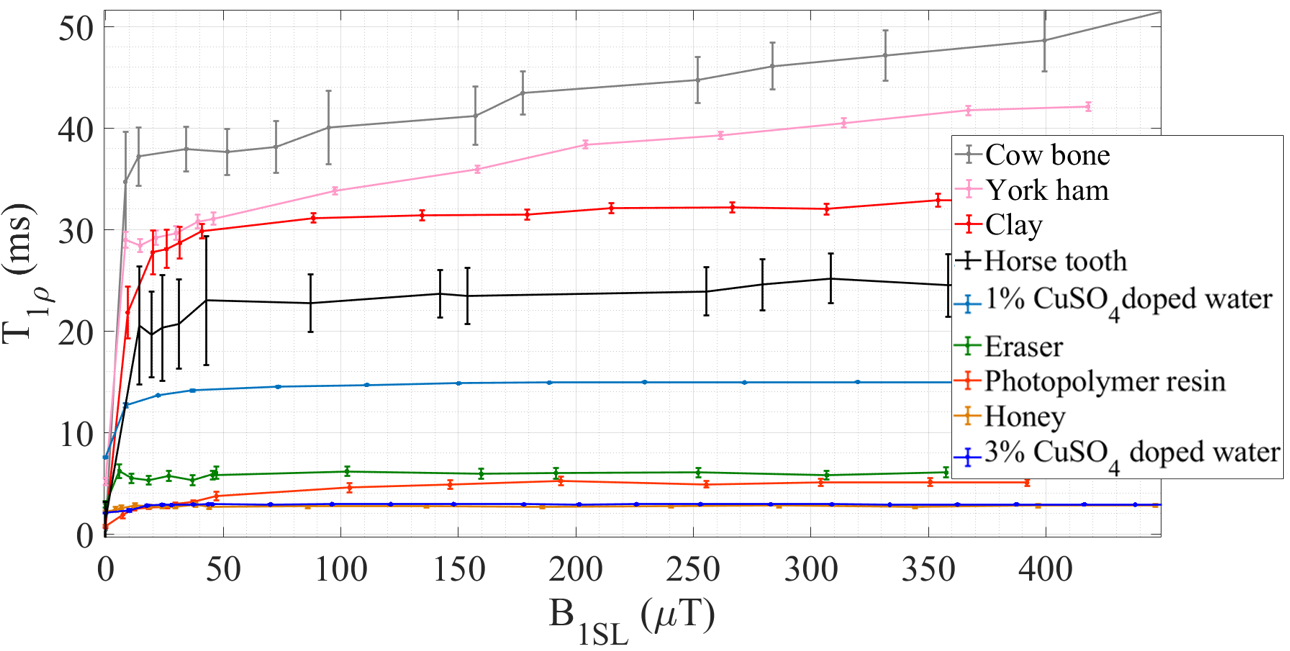}
	\caption{$T_{1\rho}$ as a function of the applied locking RF field $B_\text{1SL}$ for the samples employed throughout this work. $T_{1\rho}$ values associated to $B_\text{1SL}=0$ correspond to $T^{*}_{2}$, virtually the same as the $T_2$ values reported in Tab.\,\ref{tab:SampleRelaxationParameters} except for ham. The lines simply connect the data points and are there only for readability.}
	\label{fig:t1rhosamples}
\end{figure}

Finally, we have measured the dependence of $T_{1\rho}$ on the SL amplitude $(B_\text{1SL})$ for the samples used throughout this work (Fig.\,\ref{fig:t1rhosamples}). The main value of these measurements is that it is hard to predict $T_{1\rho}$ based on any hard/soft material distinction, and the $B_\text{1SL}$ strength that saturates $T_{1\rho}$. Thus, we found it useful to carry out these measurements before imaging a new sample to optimize the DiSLoP and PreSLoP sequence parameters. Every data point in Fig.\,\ref{fig:t1rhosamples} is determined from sequences consisting only of the excitation and spin-locking blocks, where we sweep the duration $t_\text{SL}$ of the latter. For every value of $t_\text{SL}$, we use a single data point\footnote{In fact, we read in a complete free induction decay curve after each SL pulse and we select a data point around 1\,ms after the beginning of the acquisition. In this way we ensure that ringing in the RF Tx line due to the prolonged SL pulses has faded away.}, and we fit an exponential model to determine $T_{1\rho}$ from the data set for a given $B_\text{1SL}$. This procedure is repeated for all the $B_\text{1SL}$ values in the plot, where the lines are included merely to guide the eye. As expected, for low $B_\text{1SL}$ the decay in the SL rotating frame $T_{1\rho}$ tends to $T_2^*$. For higher $B_\text{1SL}$ it can increase to a significant fraction of $T_1$, which is one of the key advantages of the proposed slice selection method. We observe $T_{1\rho}< T_1$ in all samples except for bone, consistent with expectations ($T_{1\rho}$ cannot get much longer than $T_1$ for soft tissues, see \cite{Kelly1992}). Also, note that there are two $T_1$ and $T_2$ values for bone and dental samples in Tab.\,\ref{tab:SampleRelaxationParameters}, because they consist of two types of tissues, but we only show the average $T_{1\rho}(B_\text{1SL})$ curve in Fig.\,\ref{fig:t1rhosamples} to lighten the plot. Nevertheless, for the bone and dental images in Figs.\,\ref{fig:bone} and \ref{fig:horsetooth}, we give both $T_{1\rho}$ values for each sample at the corresponding $B_\text{1SL}$.

\subsection{DiSLoP and PreSLoP for structured phantoms}
\label{sec:phantoms}

\begin{figure*}
	\centering
	\includegraphics[width=1.8\columnwidth]{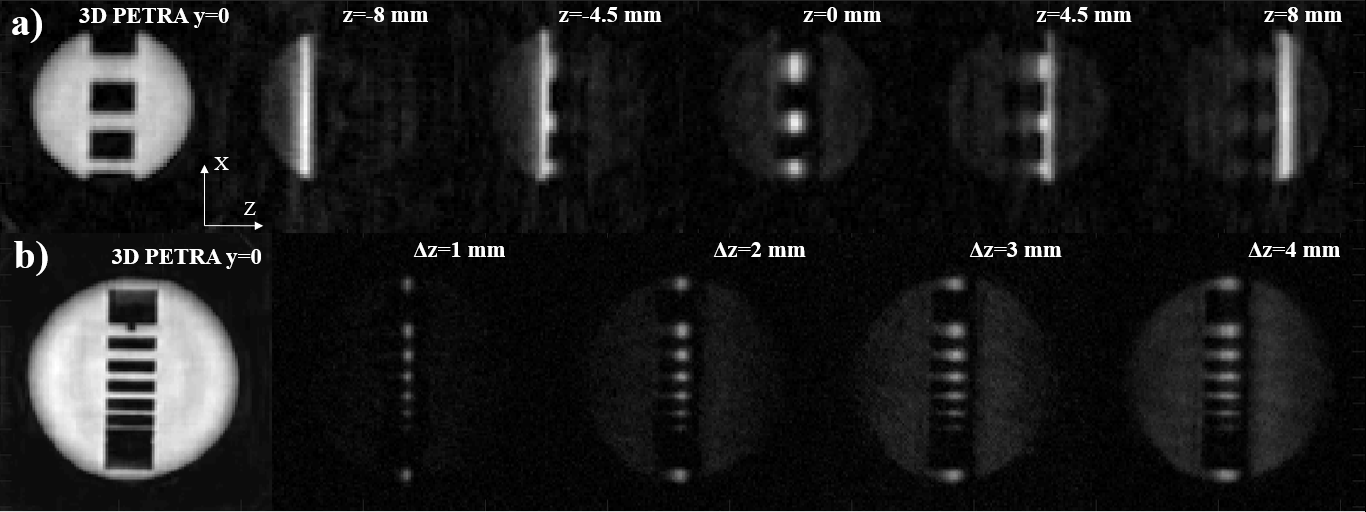}
	\caption{Control over slice selection for two PLA structured phantoms filled with 3\% CuSO$_{4}$ doped water. Leftmost images correspond to PETRA, while the rest are obtained with 3D-DiSLoP, with a) slice positions $z=-8$, -4.5, 0, 4.5 and 8 mm, and b) slice thicknesses $\Delta z=$1, 2, 3, 4 mm.}
	\label{fig:3DDiSLoPcontrol}
\end{figure*}

In this section we test the imaging capabilities of DiSLoP and PreSLoP with short and ultra-short $T_{2}$ samples in structured (rather than homogeneous) phantoms. To demonstrate control over the \emph{slice position}, the phantom in Fig.\,\ref{fig:3DDiSLoPcontrol}a was filled with 3\,\% CuSO$_{4}$ doped water ($T_2\approx 2.5$\,ms). The top row includes the $y=0$ slice from a 3D-PETRA acquisition (top left), followed by 3D-DiSLoP reconstructions where slice selection is along the $z$ direction, namely at $z=-8$, -4.5, 0, 4.5 and 8\,mm (corresponding to detunings of -20.4, -11.5, 0, 11.5, 20.4\,kHz with respect to the bare Larmor frequency). Here we use $B_\text{1SL} \approx \SI{90}{\micro T}$, $g_\text{SL}\approx 60$\,mT/m, $t_\text{SL} = \SI{800}{\micro s}$ and $\Delta z \approx 3$\,mm. To show control over the \emph{slice thickness} (Fig.\,\ref{fig:3DDiSLoPcontrol}b) we use a different phantom, but also filled with 3\,\% CuSO$_{4}$ doped water. The sequence parameters in this case are: $g_\text{SL}\approx 60$\,mT/m and $t_\text{SL} = \SI{1}{ms}$, with $B_\text{1SL}$ values ranging from 1\,mm (\SI{30}{\micro T}) to 4\,mm (\SI{120}{\micro T}) in steps of 1\,mm. As expected, the Lorentzian spatial profile (consequence of square SL pulses) becomes more evident for thicker slices. From this perspective, thinner slices are preferable as long as the SNR suffices. For this reason, we have tried to balance slice thickness and scan durations, so all the images presented in Secs.\,\ref{sec:phantoms} and \ref{sec:biologicaltissues} have a fixed slice thickness of $\approx 3$\,mm.

\begin{figure}
	\centering
	\includegraphics[width=1\columnwidth]{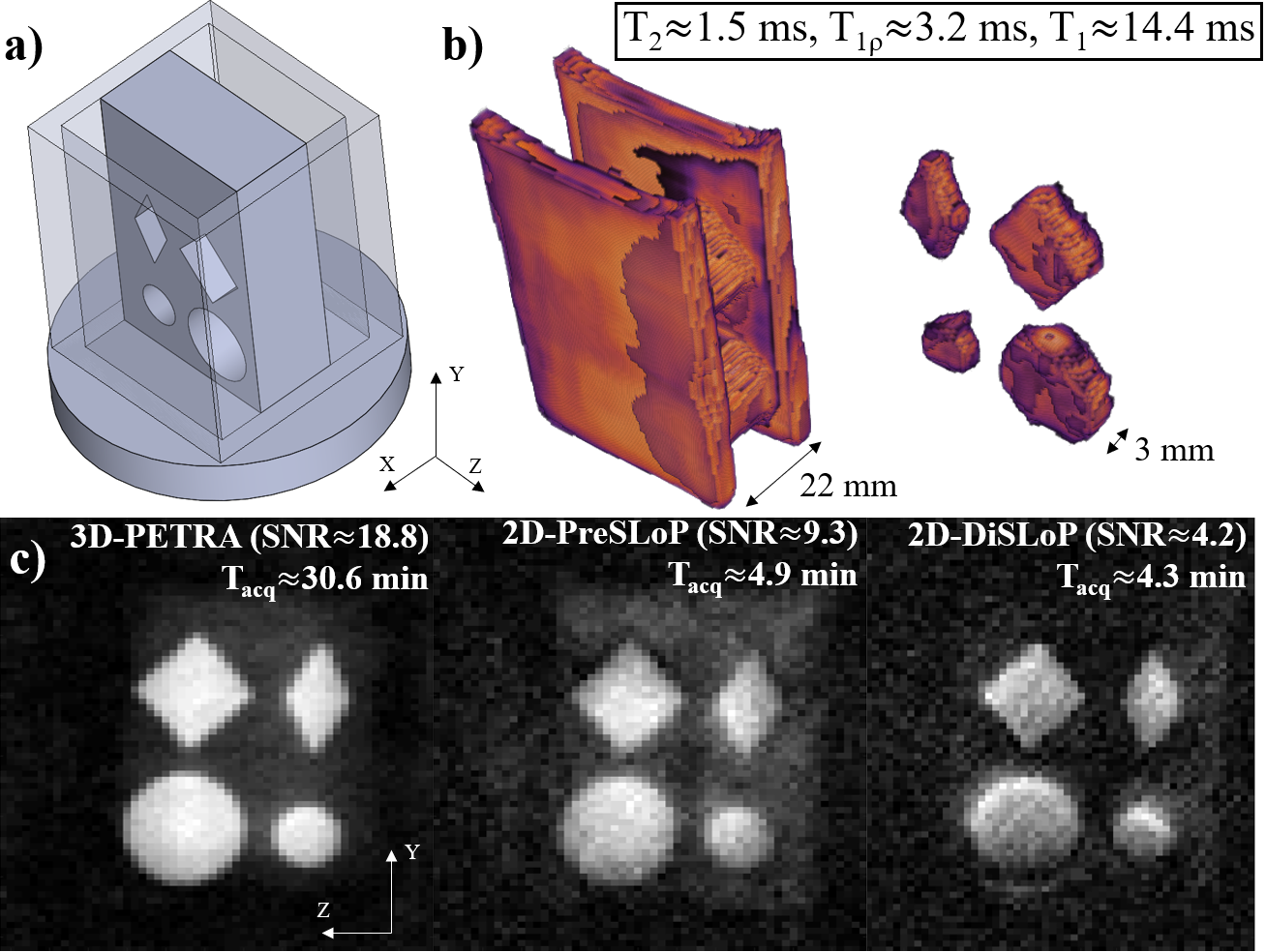}
	\caption{Performance of DiSLoP and PreSLoP for a short $T_2$ sample (honey). a) CAD image of the PLA holder for the honey. b) 3D reconstructions with PETRA (left) and 3D-PreSLoP (right), where the slice selection effect becomes evident. c) 2D images with PETRA (left) with slice resolution $\approx3$\,mm, and PreSLoP (middle) and DiSLoP (right) with $\Delta z\approx3$\,mm.}
	\label{fig:honeysample}
\end{figure}

\begin{figure}
	\centering
	\includegraphics[width=1\columnwidth]{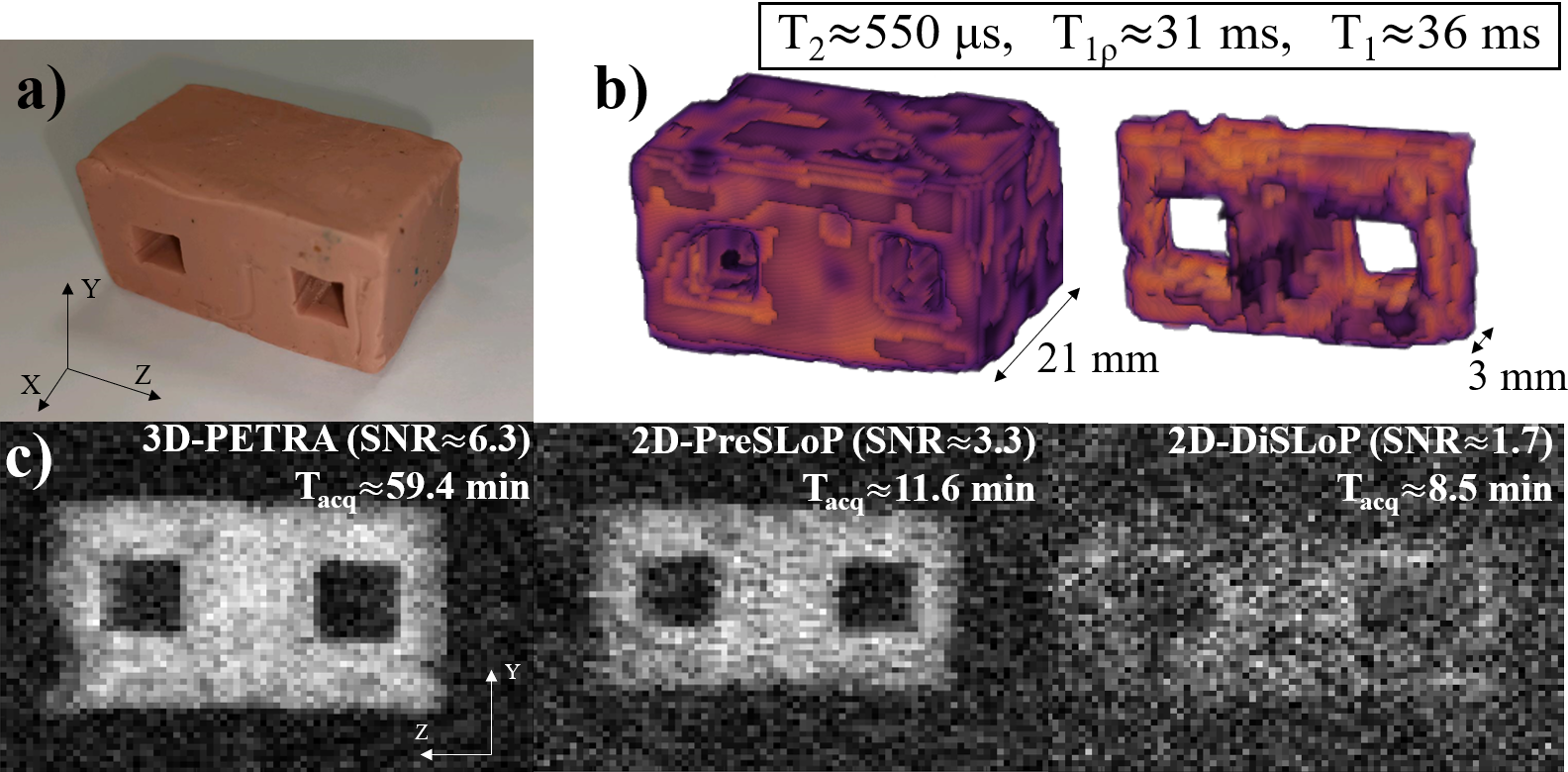}
	\caption{Performance of DiSLoP and PreSLoP for an ultra-short $T_2$ sample (clay). a) Image of the sample. b) 3D reconstructions with PETRA (left) and 3D-PreSLoP (right), where the slice selection effect becomes evident. c) 2D images with PETRA (left) with slice resolution $\approx3$\,mm, and PreSLoP (middle) and DiSLoP (right) with $\Delta z\approx3$\,mm.}
	\label{fig:clayphantom}
\end{figure}

Next we test the performance of DiSLoP and PreSLoP as compared to standard PETRA for short and ultra-short $T_{2}^*$ 2D MRI. From this point on, we plot PETRA vs 3D-PreSLoP in the upper row, and 2D images in the lower row, with PETRA on the left, 2D-PreSLoP in the middle and 2D-DiSLoP on the right.

In Figs.\,\ref{fig:honeysample} and \ref{fig:clayphantom} we have used, respectively, a PLA 3D-printed phantom filled with honey ($T_2\approx1.5$\,ms), and a phantom made of modeling clay ($T_2 \approx \SI{550}{\micro s}$). We have gathered all sequence parameters in Tab.\,\ref{tab:imageParameters} for quick reference. Figure\,\ref{fig:honeysample}(b) shows 3D reconstruction of the honey sample from a standard PETRA sequence (left, where the full sample is excited), and 3D DiSLoP (right, where the central slice has been selected with SL). In Fig.\,\ref{fig:honeysample}c) we compare 2D images of the central slice ($z=0$, $\Delta z\approx 3$\,mm): left) 3D PETRA sequence, middle) 2D-PreSLoP and right) 2D-DiSLoP (with gradient ramps of \SI{100}{\micro s}, so data acquisition starts \SI{400}{\micro s} after phase locking). Both DiSLoP and PreSLoP are capable of obtaining enough SNR to distinguish the internal phantom structures. The PreSLoP image is slightly contaminated due to imperfections in the storage and reexcitation pulses. Also, DiSLoP reveals slight shape distorsions probably due to Eddy currents created by the fast gradient transitions in the rephasing block. To avoid this, the DiSLoP acquisition could start much later, but this would lead to an increased $T_2^*$ decay. The SNR in the DiSLoP image is much lower than that of PreSLoP, since DiSLoP is subject to $T_2^*$ decay during gradient switching. On the other hand, the PETRA image has twice as much SNR as PreSLoP, due to the increased scan time $t_\text{scan}$ (i.e. averaging). To double the SNR with PreSLoP we would require $t_{\text{scan}}\approx 20$\,min ($\times 1.5$ faster than PETRA). In Sec.\,\ref{sec:discussion} we discuss in detail the expected performance of PETRA, DiSLoP and PreSLoP in terms of $t_\text{scan}$ and SNR.

\begin{table*}[t]
	\centering
	\caption{Image acquisition parameters. ``NA'' and ``NR'' stand for ``not applicable'' and ``not relevant'' respectively.}
	\label{tab:imageParameters}
	\resizebox{\textwidth}{!}{
\begin{tabular}{|c|c|c|c|c|c|c|c|c|c|c|c|c|c|}
\hline
Image        & Sequence & \begin{tabular}[c]{@{}c@{}}spin-locking\\ parameters \\ \{$B_\text{1SL}$, $g_\text{SL}$, $t_\text{SL}$\} \\ \{$\mu$T, mT/m, $\mu$s\} \end{tabular} & \begin{tabular}[c]{@{}c@{}}Nyquist \\ undersampling\end{tabular} & \begin{tabular}[c]{@{}c@{}}FOV\\ (mm$^3$)\end{tabular} & \begin{tabular}[c]{@{}c@{}}Pixel \\ size\\ (mm$^3$)\end{tabular} & \begin{tabular}[c]{@{}c@{}}Dead\\ time (us) /\\ Acquisition\\ time (us)\end{tabular} & \begin{tabular}[c]{@{}c@{}}Bandwidth\\ (kHz)\end{tabular} & \begin{tabular}[c]{@{}c@{}}TR\\ (ms)\end{tabular} & \begin{tabular}[c]{@{}c@{}}Radial\\ spokes \end{tabular} & \begin{tabular}[c]{@{}c@{}}Single\\ points\end{tabular} & Averages & \begin{tabular}[c]{@{}c@{}}Scan\\ time\\ (min)\end{tabular} & \begin{tabular}[c]{@{}c@{}}SNR \end{tabular} \\ \hline

Fig.\ref{fig:honeysample}(b)Left    & PETRA    & NA                                                      & 2                                                         & 28$\times$36$\times$28                             & 0.5$\times$0.5$\times$1                            & 85 / 1000               &36                                                              &  75                                                & 3276                                                   & 56                                                     & 10    & 41.65   & NR 
                                                       \\ \hline
Fig.\ref{fig:honeysample}(b)Right   & 3D-DiSLoP    & \{60, 40, 750 \}                                                       & 2                                                         & 28$\times$36$\times$28                             & 0.5$\times$0.5$\times$1                            & NA / 1000               &36                                                              &  75                                                & 3276                                                   & 16                                                     & 10    & 41.15    & NR  
                                                        \\ \hline
Fig.\ref{fig:honeysample}(c)Left  & PETRA    & NA                                                       & 2                                                         & 28$\times$36$\times$30                             & 0.5$\times$0.5$\times$3                            & 85 / 1000               &36                                                              &  75                                                & 1222                                                   & 0                                                    & 20   & 30.55    & 18.7   
                                                       \\ \hline
Fig.\ref{fig:honeysample}(c)Middle   & 2D-PreSLoP    & \{60, 40, 750 \}                                                       & 2                                                        & 28$\times$36                             & 0.5$\times$0.5$\times$3                            & 85 / 1000               &36                                                              &  75                                                & 160                                                   & 36                                                    & 20   & 4.90     & 9.3  
                                                       \\ \hline
Fig.\ref{fig:honeysample}(c)Right    & 2D-DiSLoP    & \{60, 40, 750 \}                                                       & 2                                                        & 28$\times$36                             & 0.5$\times$0.5$\times$3                            & NA / 1000               &36                                                              &  75                                                & 160                                                   & 12                                                    & 20   & 4.30      & 4.2                                                        \\ \noalign{\hrule height 2pt}

Fig.\ref{fig:clayphantom}(b)Left    & PETRA    & NA                                                      & 8                                                         & 40$\times$30$\times$30                             & 0.5$\times$0.5$\times$0.5                            & 80 / 750               &53.3                                                              &  50                                                & 1926                                                   & 312                                                     & 25    & 46.61     & NR                                                        \\ \hline
  
Fig.\ref{fig:clayphantom}(b)Right    & 3D-PreSLoP    & \{90, 60, 600 \}                                                       & 8                                                         & 40$\times$30$\times$30                             & 0.5$\times$0.5$\times$0.5                            & 80 / 750               &53.3                                                              &  50                                                & 1926                                                   & 312                                                     & 25    & 46.61    & NR                                                         \\ \hline  

Fig.\ref{fig:clayphantom}(c)Left    & PETRA    & NA                                                      & 2                                                         & 40$\times$30$\times$30                             & 0.5$\times$0.5$\times$3                            & 90 / 600               &66.6                                                              &  120                                                & 1358                                                   & 128                                                     & 20    & 59.40    & 6.3                                                      \\ \hline

Fig.\ref{fig:clayphantom}(c)Middle    & 2D-PreSLoP    & \{90, 60, 600 \}                                                      & 2                                                         & 40$\times$30                             & 0.5$\times$0.5$\times$3                            & 90 / 600               &66.6                                                              &  120                                                & 178                                                   & 112                                                     & 20    & 11.60    & 3.3                                                          \\ \hline

Fig.\ref{fig:clayphantom}(c)Right    & 2D-DiSLoP    & \{90, 60, 600 \}                                                       & 2                                                         & 40$\times$30                             & 0.5$\times$0.5$\times$3                            & NA / 600               &66.6                                                              &  120                                                & 178                                                   & 36                                                     & 20    & 8.50      & 1.8    
                                                    \\ \noalign{\hrule height 2pt}

Fig.\ref{fig:bone}(b)Left    & PETRA    & NA                                                      & 8                                                         & 46$\times$34$\times$30                             & 1$\times$1$\times$1                            & 80 / 600               &38.3                                                              &  75                                                & 572                                                   & 96                                                     & 25    & 20.88     & NR                                                        \\ \hline

Fig.\ref{fig:bone}(b)Right    & 3D-PreSLoP    & \{150, 100, 100 \}                                                       & 8                                                         & 46$\times$34$\times$30                             & 1$\times$1$\times$1                            & 80 / 600               &38.3                                                              &  75                                                & 572                                                   & 96                                                     & 25    & 20.88     & NR                                                        \\ \hline  

Fig.\ref{fig:bone}(c)Left    & PETRA    & NA                                                       & 2                                                         & 46$\times$34$\times$30                             & 0.5$\times$0.5$\times$3                            & 90 / 600               &76.6                                                              &  75                                                & 1556                                                   & 152                                                     & 15    & 32.03      & 3.8                                                         \\ \hline  

Fig.\ref{fig:bone}(c)Middle    & 2D-PreSLoP    & \{150, 100, 100 \}                                                       & 2                                                         & 46$\times$34                             & 0.5$\times$0.5$\times$3                            & 90 / 600               &76.6                                                              &  75                                                & 204                                                   & 136                                                     & 15    & 6.34      & 3.6                                                         \\ \hline 

Fig.\ref{fig:bone}(c)Right    & 2D-DiSLoP    & \{150, 100, 100 \}                                                       & 2                                                         & 46$\times$34                             & 0.5$\times$0.5$\times$3                            & NA / 600               &76.6                                                              &  75                                                & 204                                                   & 44                                                     & 15    & 4.65      & 1.0                                                         \\ \noalign{\hrule height 2pt}

Fig.\ref{fig:horsetooth}(b)Left    & PETRA    & NA                                                      & 8                                                         & 38$\times$90$\times$21                             & 1$\times$1$\times$1                            & 110 / 600               &75                                                              &  25                                                & 764                                                   & 288                                                     & 375    & 164.37      & NR                                                       \\ \hline  

Fig.\ref{fig:horsetooth}(b)Right    & 3D-PreSLoP    & \{90, 60, 190 \}                                                       & 8                                                         & 38$\times$90$\times$21                             & 1$\times$1$\times$1                            & 110 / 600               &75                                                              &  25                                                & 764                                                   & 288                                                     & 375    & 164.37     & NR                                                        \\ \hline

Fig.\ref{fig:horsetooth}(c)Left    & PETRA    & NA                                                       & 2                                                         & 38$\times$90$\times$21                             & 1$\times$1$\times$3                            & 110 / 600               &75                                                              &  25                                                & 1010                                                   & 112                                                     & 150    & 70.12     & 6.2                                                          \\ \hline

Fig.\ref{fig:horsetooth}(c)Middle    & 2D-PreSLoP    & \{90, 60, 190 \}                                                       & 2                                                         & 38$\times$90                             & 1$\times$1$\times$3                            & 110 / 600               &75                                                              &  25                                                & 200                                                   & 112                                                     & 150    & 19.37     & 4.7                                                          \\ \hline

Fig.\ref{fig:horsetooth}(c)Right    & 2D-DiSLoP    & \{90, 60, 190 \}                                                       & 2                                                         & 38$\times$90                             & 1$\times$1$\times$3                            & NA / 600               &75                                                              &  25                                                & 200                                                   & 24                                                     & 150    & 14.00      & 1.2                                                         \\ \noalign{\hrule height 2pt}
\end{tabular}}
\end{table*}

We have followed the exact same procedure to analyze the clay phantom reconstructions. The clay used exhibits an ultra-short $T_2 \approx \SI{550}{\micro s}$. In Fig.\,\ref{fig:clayphantom} we now observe a significantly lower SNR in DiSLoP than in PreSLoP, even for similar $t_{\text{scan}}$ (see Tab.\,\ref{tab:imageParameters}). Note that DiSLoP would need $\times 2.5$ averages (i.e. $\approx 21$\,min) to reach the SNR in the PreSLoP reconstruction. Similarly, to obtain an SNR$=6.33$ (that of PETRA) for PreSLoP we would need $t_{\text{scan}}\approx 44$\,min ($\times 1.36$ faster than PETRA).

\subsection{Hard biological tissues}
\label{sec:biologicaltissues}

\begin{figure}[h!]
	\centering
	\includegraphics[width=1\columnwidth]{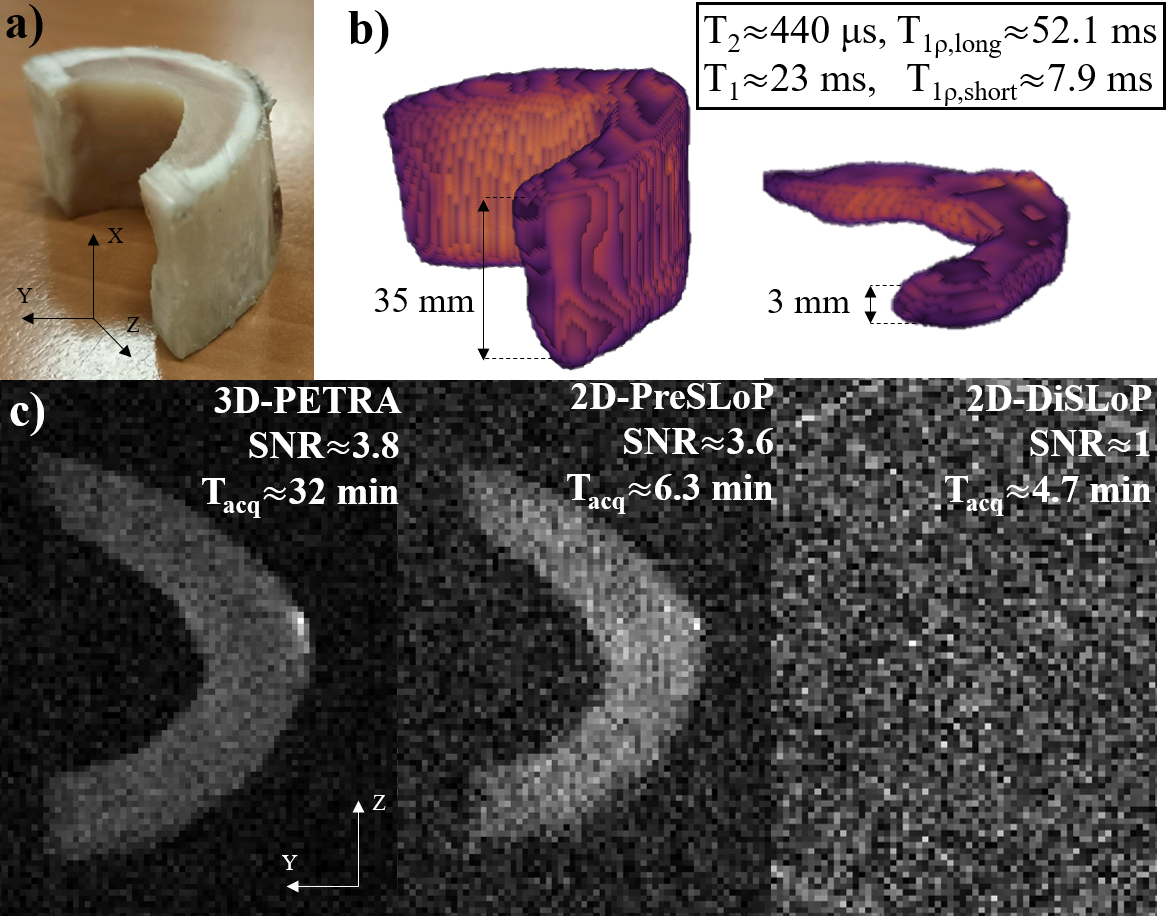}
	\caption{Performance of DiSLoP and PreSLoP for an ultra-short $T_2$ biological tissue (cortical bone). a) Photograph of bovine femur sample. b) 3D PETRA (left) and 3D-PreSLoP (right). c) 2D images for PETRA (left) with slice resolution $\approx 3$\,mm, PreSLoP (middle) and DiSLoP (right) with $\Delta z\approx 3$\,mm. We provide $T_{1\rho}$ for both tissues, and the average $T_1$ and $T_2$ values (see Tab.\,\ref{tab:SampleRelaxationParameters}).}
	\label{fig:bone}
\end{figure}

\begin{figure}[h!]
	\centering
	\includegraphics[width=0.85\columnwidth]{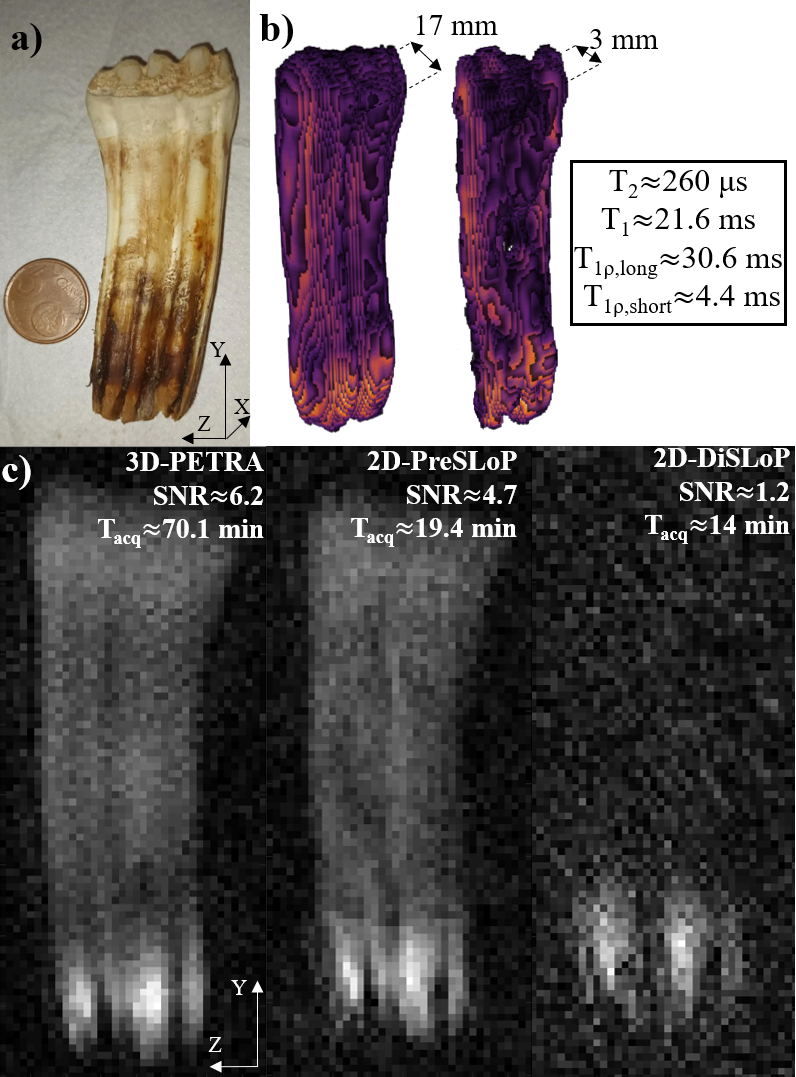}
	\caption{Performance of DiSLoP and PreSLoP for an ultra-short $T_2$ biological tissue (horse tooth). a) Photograph of the sample. b) 3D PETRA (left) and 3D-PreSLoP (right). c) 2D images for PETRA (left) with slice resolution $\approx 3$\,mm, PreSLoP (middle) and DiSLoP (right) with $\Delta z \approx 3$\,mm. We provide $T_{1\rho}$ for both tissues, and the average $T_1$ and $T_2$ values (see Tab.\,\ref{tab:SampleRelaxationParameters}).}
	\label{fig:horsetooth}
\end{figure}

Finally, following the same procedure as in previous figures, we present images of \textit{ex vivo} hard biological tissues obtained with DiSLoP and PreSLoP and compare them with PETRA in terms of SNR and scan time for the same slice and in-plane resolution.

The images in Fig.\,\ref{fig:bone} are from a piece of cortical bone from a bovine femur ($T_2 \approx \SI{440}{\micro s}$), obtained from a local butcher and stripped off of the surrounding soft tissues. Figure\,\ref{fig:bone}b shows 3D reconstructions using both a standard PETRA sequence, where the full sample is excited and imaged, and 3D-PreSLoP. Details on the sequence parameters are provided in Tab.\,\ref{tab:imageParameters}. Given the ultra-short $T_{2}$ values in the sample, PreSLoP performs significantly better than DiSLoP, where the resulting image is dominated by noise despite a very short rephasing block of only \SI{400}{\micro s}. Regarding SNR, PreSLoP and PETRA yield very similar results, but the acquisition with the former is $\times 5$ faster. 

Next, we show results for the shortest $T_2$ sample used in this work: an \textit{ex vivo} horse tooth stripped off of soft tissues and dehydrated to remove water residues in internal cavities. The average $T_2$ for this sample is only around \SI{260}{\micro s}. Again, DiSLoP fails to form an image showing the hardest tissues. Only the lower part of the tooth, which is significantly softer ($T_2 \approx 6$\,ms), is visible. For a similar scan time PreSLoP shows a much better image. In order to reach the SNR of the PETRA acquisition, PreSLoP would need $\approx 35$\,min, i.e. a factor $\times 2$ faster than PETRA.

\section{Discussion}
\label{sec:discussion}
From the images in Figs.\,\ref{fig:honeysample}-\ref{fig:horsetooth} we conclude that PreSLoP allows for slice-selective ZTE of ultra-short $T_2$ samples, which was the main goal of this work. Admittedly, there is some contamination in the PreSLoP reconstruction of the honey sample (Fig.\,\ref{fig:honeysample}), but this should be solvable with improved RF hardware (see Sec.\,\ref{sec:control}). DiSLoP performs reasonably well with samples where the measured $T_2$ is $\gtrsim 1$\,ms, but otherwise fails. This highlights the critical importance of the preservation block for hard biological tissues and solid-state matter. 

The devised sequences can be modified in at least two ways: i) for 2D-DiSLoP the encoding gradient can be ramped up during the rephasing block (with an inverted triangle as shown for the slice selection gradient) so that acquisitions start at $k=0$, thus eliminating the dead time (i.e. the $k$-space gap) for this sequence, given that Eddy currents can be calibrated for each amplitude of the encoding gradient; and ii) the final spin excitation (flip angle $\theta$) in 2D-PreSLoP could be pulsed before/during the encoding gradient ramp-up, as in UTE, even if this would involve a $T_2^*$ signal loss which could be intolerable for some applications.

All in all, the fact that the efficiency of each sequence depends on detailed experimental parameters (see below) and on the sample properties ($T_{1\rho}$ and $T_2$) leaves much room for sequence optimization for specific application, and shall be further explored as required.

\subsection{Efficiency comparison}

\begin{figure}[h!]
	\centering
	\includegraphics[width=1\columnwidth]{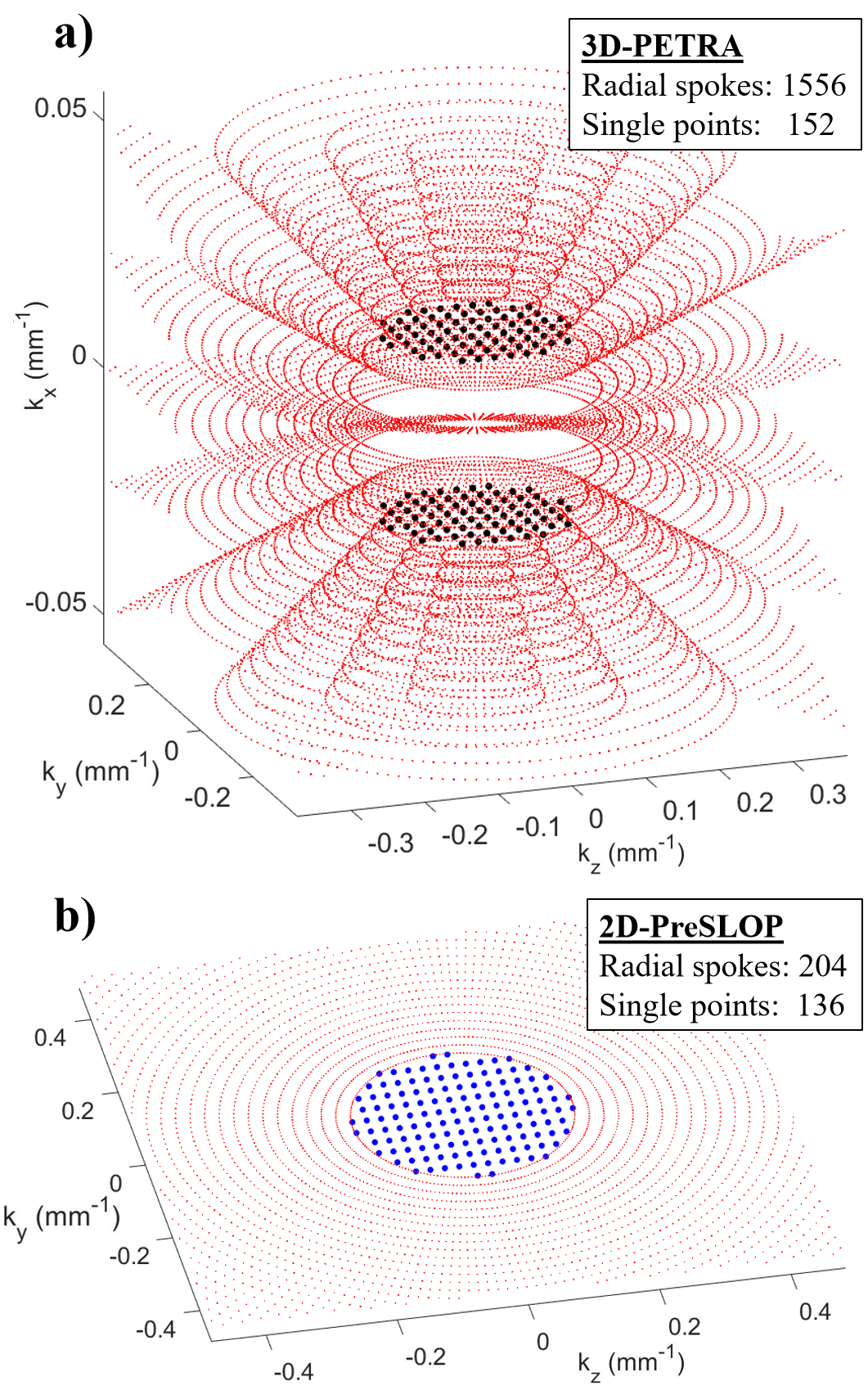}
	\caption{Zoom into the central region of $k$-space of the cow bone images presented in Fig.\,\ref{fig:bone}. a) PETRA. b) PreSLoP.}
	\label{fig:kSpaces}
\end{figure}

Here we discuss the advantage of PreSLoP with respect to DiSLoP and PETRA in terms of SNR efficiency and scan time. The underlying reasons are rather subtle and parameter-dependent, but some insight can be gained by comparing the three sampling schemes in $k$-space. Two important concepts in this regard are: i) scan time in these sequences is dominated by the number of pointwise acquisitions (radial spoke acquisitions cover many $k$-space points in a single TR); and ii) the largest contribution to the reconstruction SNR comes from the central region of $k$-space.

For illustration purposes, in Fig.\,\ref{fig:kSpaces} we plot $k$-space positions (not values) for the PETRA and 2D-PreSLoP sequences used to image the bovine femur (Fig.\,\ref{fig:bone}), highlighting the pointwise encoding regions (larger black/blue points). A priori, one may expect that the number of pointwise acquisitions in PETRA must be much larger than in 2D-PreSLoP, since the gap in the former is spherical and only a 2D surface in the latter. However, for the chosen slice direction, dead time and resolution, our PETRA implementation (for an even number of slices) requires pointwise sampling only at $k_z=\pm \delta k_z/2$. Consequently, the amount of pointwise acquisitions is similar for PETRA and 2D-PreSLoP (152 vs 136, see Tab.\,\ref{tab:imageParameters}). A slightly different configuration may have required three pointwise planes for PETRA, and 2D-PreSLoP would have scaled even more favorably. 

Compared to 2D-PreSLoP, 2D-DiSLoP requires significantly fewer points for a similar scan time (136 vs 40). This is because the $k$-space gap in DiSLoP is smaller, since the readout gradient starts to ramp when the magnetization is already transversal, whereas it is already at its full strength during the final excitation pulse in PreSLoP. 

In terms of SNR, 2D-DiSLoP always suffers $T_2^*$ decay, and it is to be expected that it performs worse than 2D-PreSLoP for the same scan time. The comparison with PETRA is less straight-forward. For instance, the SNRs of the bovine images are very similar, but PETRA takes $\times 5$ longer than 2D-PreSLoP. As mentioned above, the SNR has a heavy bias towards the number of central points in $k$-space, and the pointwise acquisitions are comparable (152 for PETRA vs 136 for 2D-PreSLoP). Hence, a PETRA sequence will need to fill a 3D $k$-space with many radial spokes, while a 2D-PreSLoP sequence will only fill a 2D $k$-space (1556 vs 204 spokes). This explains why 15 averages of PETRA and 2D-PreSLoP yield similar SNR values, but PETRA takes much longer.

To sum up, it is difficult to predict the efficiency of the different sequences for specific cases, but two rule-of-thumb guidelines could be: i) for samples with longer $T_2$ one might prefer DiSLoP to PreSLoP because the former requires a smaller pointwise region (which can be even completely avoided if Eddy currents are under control), and the contrary applies for ultra-short $T_2$ samples; and ii) 2D-PreSLoP should yield a higher SNR than PETRA per unit scan time, because PreSLoP has a 2D scaling for the pointwise region, and because PETRA's 3D filling of non-central $k$-space contributes much less to the SNR.

\subsection{Outlook}

In addition to the discussed advantages as compared to standard PETRA, the proposed 2D imaging protocols feature also several benefits with respect to the double half-pulse RF excitation for slice selection with UTE \cite{FABICH2014116}. First, our protocols are robust against Eddy currents because one can wait for gradients to stabilize and then fire the RF pulses. Second, they are faster by construction, since they do not require two acquisitions for every $k$-space line. Third, signal decay during slice selection with DiSLoP and PreSLoP is subject to $T_{1\rho}\gg T_2^*$, and thus easier to deal with than with 2D-UTE in terms of timing and hardware. For instance, well-defined rectangular slice profiles are easier to attain with SL than with a standard sinc-modulated pulse, where most SNR can get loss in the process. Finally, slice selection by SL, both with and without preservation pulse, is compatible with UTE approaches.

\begin{figure}[h]
	\centering
	\includegraphics[width=1\columnwidth]{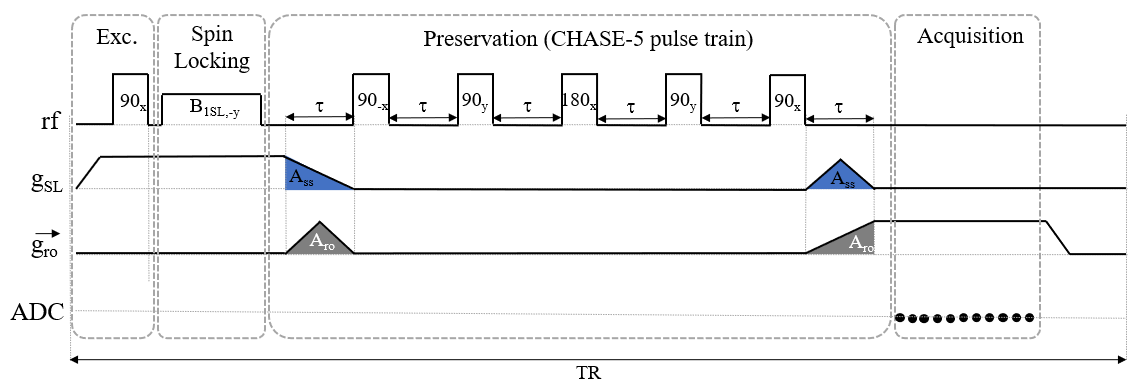}
	\caption{Possible embodiment of PreSLoP where the preservation block includes a CHASE pulse train of length 5, allowing for zero dead time before data acquisition.}
	\label{fig:CHASE}
\end{figure}

We have devised also a possible alternative to the preservation block shown in Fig.\,\ref{fig:pulseseqdiagram}, based on a CHASE-5 sequence \cite{Waeber2019}. This is shown in Fig.\,\ref{fig:CHASE} and is basically a modification of the well-known WAHUHA sequence\cite{WAHUHA} with a 180$^{\circ}$ rephasing pulse in the middle. This sequence is able to undo dephasing caused by both homogeneous (spin-spin interactions) and inhomogeneous (gradient) sources. In our case, it can be used to avoid $T_2^*$ decay during the preservation block, while the dephasing caused by ramping gradients is compensated with the waveforms shown in the figure. Note that the latter compensation does not need opposite polarity because CHASE-5's central 180$^{\circ}$ changes the effective sign of spin evolution. Because this sequence compensates gradient evolution too, the acquisition could start at $k=0$, thus eliminating the dead time central gap in $k$-space and therefore the need of pointwise encoding. This would lead to a reduced scan time at the expense of more complicated RF pulse trains and potentially increased SAR (specific absorption rate) \cite{IEC2010}.



\section{Conclusion}
\label{sec:concl}
We have presented two protocols that combine slice selection through spin-locking with Zero Time Echo 2D imaging. DiSLoP is especially suited to moderately short $T_2$ tissues (in the order of few milliseconds), and has the advantage that it can avoid TxRx switching dead time. However, for sub-millisecond $T_2$ tissues, we have presented PreSLoP, which adds a preservation pulse and is capable of 2D imaging the hardest tissues in the body: bone and teeth. PreSLoP suffers from dead time, and thus leaves a central $k$-space gap that has to be filled by Cartesian pointwise encoding, as in standard PETRA, but has a favorable SNR and 2D scaling, which can lead to significant scan time reduction. All in all, we have shown sequences that can combine superior slice selection (which suffers from $T_{1\rho}\gg T_2$ decay) with the most efficient imaging sequence available for ultra-short $T_2$ samples (ZTE) and which have enabled 2D imaging of {\it ex vivo} bone and teeth.


\section{Contributions}

Experiments were performed by JB, with analysis of experimental data by JB, JMA, FG and JA. Theoretical analysis performed by FG and JA. The paper was written by JB, FG and JA, with input from all authors. Experiments conceived by FG, JA and JMB.

\appendices

\section*{Acknowledgment}

We thank present and former members of the MRILab team for their support and participation in the design and assembly of the experimental setup, and Aloma Mayordomo Febrer, from the Clinical Veterinary Hospital of Cardenal Herrera University, for the horse tooth. This work was supported by the Ministerio de Ciencia e Innovaci\'on of Spain through research grant PID2019-111436RB-C21. Action co-financed by the European Union through the Programa Operativo del Fondo Europeo de Desarrollo Regional (FEDER) of the Comunitat Valenciana (IDIFEDER/2018/022 and IDIFEDER/2021/004). JB acknowledges support from the Innodocto program of the Agencia Valenciana de la Innovaci\'on (INNTA3/2021/17).

\section*{Ethical statement}
All animal parts were obtained from a local butcher and research was conducted following the 3R principles.

\section*{Conflict of interest}
The authors have a patent (WO2021245024A1) protecting the techniques presented in this paper and licensed to Tesoro Imaging S.L. JB is a research scientist at Tesoro Imaging S.L. JMB is a co-founder of Tesoro Imaging S.L.

\appendix
\subsection{Spin-locking equations}
The equations governing the magnetization at each infinitesimal plane in the slice selection direction ($z$), are
\begin{eqnarray}
M_x(t)&=&M_0 \cos^2\alpha +M_0 \sin^2\alpha \cos( \gamma Bt)\nonumber\\
M_y(t)&=&-M_0 \sin\alpha \sin( \gamma Bt)\\
M_z(t)&=&M_0 \sin\alpha\cos\alpha\, \left(1- \cos( \gamma Bt)\right),\nonumber
\label{eq1}
\end{eqnarray}
where decay is not included and with $B=\sqrt{B_\text{1SL}^2+(g_\text{SL}\cdot z)^2}$, $\cos\alpha=B_\text{1SL}/B$, and $\sin\alpha=g_\text{SL}\cdot z/B$, with $B=B(z)$. The main idea behind slice selection through SL is that either because of a fast $T_2$ or because the slice selection gradient is strong (thus causing a short $T_2^*$), only the average magnetization along the field $\vec{B}$ in each slice will survive. One can imagine the sample as divided in slices of width $\Delta z$, whose average magnetization is detected. For slices far from the selected one, $B(z)\simeq |g_\text{SL}\cdot z|$, thus the slice-average of e.g. $\cos \gamma Bt$ dies off as the envelop of $\text{sinc}(\gamma g_\text{SL} \Delta z\, t_\text{SL}/2)=\text{sinc}(\gamma B_\text{1SL} t_\text{SL})$. For instance, for $t_\text{SL} \gtrsim 7\pi/(2\gamma B_\text{1SL})$, a fraction $\lesssim 10$\,\% survives (Eq.\,(\ref{eq:time})). Once time-dependent terms get averaged out, a magnetization similar to $\vec{M}\simeq (M_0 \cos^2\alpha ,0,M_0 \sin\alpha\cos\alpha)$ remains, leading to an observable (i.e. transverse) magnetization
$$M_{\text{obs}}=|M_x+iM_y|\simeq M_0\cos^2\alpha=M_0\frac{B_\text{1SL}^2}{B_\text{1SL}^2+(g_\text{SL}\cdot z)^2},$$ which is Lorentzian with a full width at half maximum (FWHM) of $\Delta z\approx 2B_\text{1SL}/g_\text{SL}$ (Eq.\,(\ref{eq:thickness})). The ensemble average in a thin slice at $z=0$ has almost negligible average $M_z$ if the sample's slice is homogeneous enough, since $\sin\alpha$ is an odd function.

Of course, samples are never homogeneous, we have neglected decay, and the dynamics near the selected slice has more complicated equations, but the derived expressions give an approximate idea of expected timescales and behavior in the sequences in Fig.\,\ref{fig:pulseseqdiagram}.

\subsection{Spin-locking equations with decay}
When we include spin-spin relaxation in the rotating frame, the equations read:
\begin{eqnarray}
M_x(t)&=&M_0 \cos^2\alpha +M_0 \sin^2\alpha \cos( \gamma Bt)\, \text{e}^{-t/T_{2\rho}}\nonumber\\
M_y(t)&=&-M_0 \sin\alpha \sin (\gamma Bt)\, \text{e}^{-t/T_{2\rho}}\\
M_z(t)&=&M_0 \sin\alpha\cos\alpha\, (1- \cos( \gamma Bt)\, \text{e}^{-t/T_{2\rho}}),\nonumber
\label{eq2}
\end{eqnarray}
with $T_{2\rho}$ the transverse (dipole-dipole) decay rate in the rotating frame. This decay rate is expected to be given by \cite{Kelly1992, Wheaton2004}
\begin{equation}
 \frac{1}{T_{2\rho}}\simeq\frac{1}{2}\left(\frac{1}{T_1}+\frac{1}{T_2}\right),
\end{equation}
which for the cases of interest of spin-locking ($T_1\gg T_2$) leads to $T_{2\rho}\simeq 2T_2$. Therefore $T_{2\rho}$ accelerates slice selectivity.

If we further include spin-lattice relaxation in the rotating frame, the $M_0 \cos^2\alpha$ term in $M_x(t)$ and the $1$ term in $M_z(t)$ in Eq.\,(\ref{eq1}) suffer relaxation of magnetization towards the equilibrium Boltzmann value at a rate $T_{1\rho}$.  If we wait long enough ($t_\text{SL}\gg T_{1\rho} > T_{2\rho}$) the final magnetization is $\vec{M}\simeq (M_0\cos\alpha,0,M_0\sin\alpha)$, that is, the transverse magnetization goes now with $\cos\alpha$ instead of $\cos^2\alpha$, leading to a broader slice profile.

\subsection{Advanced spin-locking}
Selection of a squared-shape slice profile with a sinc-modulate $B_\text{1SL}(t)$ RF locking-pulse can be argued within the small-flip angle approximation \cite{HandbookPulses}, which works very well for flip-angles below $30^\circ$ and approximately well for flip-angles until $90^\circ$. In this sense, spin-locking simply locks resonantly only the spin-frequencies ($\gamma g_\text{SL}\cdot z$) which fall under the Fourier transform of $B_\text{1SL}(t)$ as in standard slice selection. In Fig.\,\ref{fig:SquareSL} we see that indeed this approximation works well and an approximately squared-shaped spin-selection profile is achieved.

By the same argument, multi-slice selection occurs when the time-profile of $B_\text{1SL}(t)$ has a Fourier transform with two peaks, which is achieved by a double-tone $B_\text{1SL}(t)=B_1 \cos(\gamma (B_0+g_\text{SL}\, z_1)t)+B_1 \cos(\gamma (B_0+g_\text{SL}\, z_2)t)$. To see how this works in detail we can move to a frame rotating at $\gamma B_0+g_\text{SL}\, z_1$, with a field:
\begin{eqnarray}
 \vec{B}&=&g_\text{SL}(z-z_1)\hat{u}_z+\frac{B_1}{2}\hat{u}_x+\nonumber\\
 &&\frac{B_2}{2}(\hat{u}_x\cos[(\omega_1-\omega_2)t]+\hat{u}_y\sin[(\omega_1-\omega_2)t]).
\end{eqnarray}
For spins near $z_1$ and when $\omega_1-\omega_2=\gamma g_\text{SL} (z_1-z_2)\gg \gamma B_1$, the oscillatory terms are averaged out in the timescale relevant to spin-locking ($2\pi/\gamma B_1$), resulting in $\vec{B}=g_\text{SL}(z-z_1)\hat{u}_z+\frac{B_1}{2}\hat{u}_x$. This locks the spins near $z_1$ and dephases the rest. Repeating the argument in the frame rotating at $\gamma B_0+g_\text{SL}\, z_2$, we obtain $\vec{B}=g_\text{SL}(z-z_2)\hat{u}_z+\frac{B_1}{2}\hat{u}_x$ which spin-locks near $z_2$. As an example, consider a slice selection gradient of $70$\,mT/m and a spin-locking amplitude that selects a slice with $d=1$\,mm (so that $B_1=g_\text{SL}\, d/2$). The characteristic evolution is given mainly by $\gamma B_1\approx 2\pi\cdot 1.5$\,kHz, while oscillatory terms evolve at $\omega_2-\omega_1\approx 2\pi\cdot 30$\,kHz for double-tone spin-locking with $|z_1-z_0|=1$\,cm, and thus average to zero.

\subsection{SNR estimations}

\begin{figure}[h!]
	\centering
	\includegraphics[width=0.8\columnwidth]{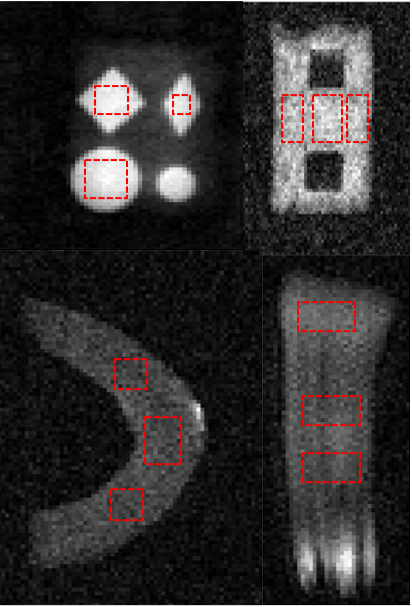}
	\caption{Regions of interest selected for SNR estimations in Figs.\,\ref{fig:honeysample}-\ref{fig:horsetooth}.}
	\label{fig:ROIs}
\end{figure}

The SNR values given in Figs.\,\ref{fig:honeysample}-\ref{fig:horsetooth} are estimated by averaging the signal strength in three bright regions (inside each sample), divided by the average signal strength in a dark region (background). Figure\,\ref{fig:ROIs} shows the regions of interest selected for each sample: honey (top left), clay (top right), bone (bottom left) and tooth (bottom right).

\subsection{Effect of gradient polarity on PreSLoP performance}

\begin{figure}[h!]
	\centering
	\includegraphics[width=1\columnwidth]{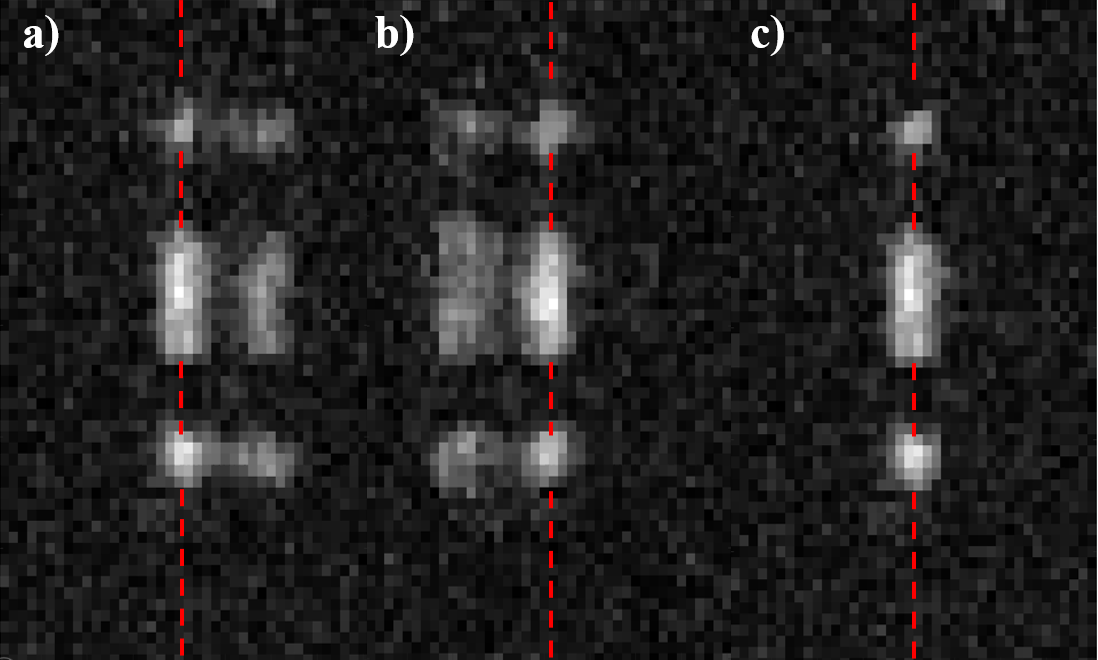}
	\caption{Effect of the $g_\text{SL}$ polarity over slice profile definition obtained with 3D-PreSLoP and the clay phantom in Fig.\,\ref{fig:clayphantom}: a) positive $g_\text{SL}$ in all repetitions; b)  negative $g_\text{SL}$ in all repetitions; c) alternating $g_\text{SL}$ polarity between repetitions. Red dotted lines mark the $x=0$ position in each image, where the slice is nominally selected.}
	\label{fig:polarity}
\end{figure}

PreSLoP is subject to distortions in the slice profile for short repetition times, presumably due to Eddy currents inducing drifts on $B_0$. Inverting the polarity of the spin-locking gradient ($g_\text{SL}$) for each radial acquisition mitigates this effect, because the contributions average out to a large extent. For illustration purposes, Fig.\,\ref{fig:polarity} shows the effect of: i) employing a single polarity on a 3D-PreSLoP (positive on the left image, negative on the middle image), where artifacts are clearly visible; and ii) alternating polarity between repetitions, where they are largely suppressed. These images have been taken with the clay phantom in Fig.\,\ref{fig:clayphantom}.

\ifCLASSOPTIONcaptionsoff
  \newpage
\fi


\end{document}